\documentclass[preprint,eqsecnum,preprintnumbers,nofootinbib,byrevtex,prd,aps,showpacs,showkeys,groupedaddress,floatfix]{revtex4}
\usepackage{bm}
\usepackage{graphics}
\usepackage{graphicx}
\usepackage{amssymb}
\usepackage{amsmath}

\begin{document}

\title{\textbf{Electromagnetic interaction vertex of $\Delta$ baryons in
hard-wall AdS/QCD model}}
\author{Shahin Mamedov ${}^{1,2}$}
\email{sh_mamedov@bsu.az}
\affiliation{${\ }^{(1)}$ Institute for Physical Problems, Baku State University,
Z.Khalilov str. 23, Baku, AZ-1148, Azerbaijan \\
${\ }^{(2)}$Department of Physical Engineering, Ankara University, Tandogan
06100, Ankara, Turkey}
\date{\today }

\begin{abstract}
We consider an interaction of a spin-3/2 field with the electromagnetic field
in the framework of the hard-wall model of AdS/QCD. We write Lagrangian for
this interaction including all kinds interaction terms in the bulk of AdS$_{5}$ 
space and present the scattering matrix element in integrals over the
fifth coordinate. Comparing the current matrix element obtained in the
boundary of this space with the one known from field theory, we find the
vertex function coefficients for the $\gamma^{*}\Delta\Delta$ interaction
vertex. 
As an example, we apply the obtained coefficients to the computation of the 
charge form factor $G_{E0}$ for the $\Delta^{+}$ baryon and compare the result with
the one obtained in the field theory.
\end{abstract}

\pacs{14.20.Gk, 13.40.Gp, 11.25.Tq, 11.10.Kk}
\maketitle

\section{Introduction}

The idea of the AdS/CFT (Anti de-Sitter/Conformal Field Theory)
correspondence~\cite{1} (or gravity-gauge duality) is successfully used for
solving problems in different areas of theoretical physics. Originally
formulated for a super Yang-Mills theory this duality connects fields in the
bulk of the AdS space with current correlators on the boundary of this space
which then are corresponded to the ones in the field theory in flat
space-time with less dimension~\cite{2}. In application to QCD the duality
idea is valuable for solving low-energy problems, when usual perturbation
theory does not work. Though QCD is not conformal theory, there are some QCD
models based on this duality, such as holographic QCD and AdS/QCD models. In
order to build a model of AdS/QCD one should break the conformal symmetry.
This can be done either by imposing boundary condition on a solution to
equation of motion over extra dimension or by warping the metric of the AdS
space. In the first case, the boundary condition cuts a slice of the AdS
space and gives mass spectrum, which then is corresponded to the mass
spectrum of observed particles. Such a model is called a hard-wall model~
\cite{3,4,5} and is successfully applied to calculation of physical
quantities of mesons and baryons~\cite{6,7,8,9,10,11,12,18,19,20,21}. In the
second case, which is called a soft-wall model, there are no sharp
boundaries and propagation of a particle at long distances of extra
dimension is suppressed by introduction of a 
dilaton field~\cite{22,23,24,25,26}.

The hard-wall model was extended in~\cite{11} by inclusion of spin-3/2
fields and the model was applied to calculations of the meson-baryon
transition couplings and transition form factors for interactions of $\Delta$
baryons with nucleons, $\pi $ and $\rho $ mesons. Here we include
interactions of spin-3/2 fields with a photon field into the hard-wall model
and consider $\gamma ^{\ast }\Delta \Delta $ interaction vertex within this
model. In Section~\ref{sec2} we briefly describe the hard-wall model with
the spin-3/2 field introduced in~\cite{11}. In Section~\ref{sec3}, we
present some relevant formulas of the isotopic structure for $\Delta$
baryons and expressions of multipole form factors of these particles. We
construct in the bulk of the AdS$_{5}$ space an interaction Lagrangian for
the spin-3/2 field with a gauge field, which includes all possible
interactions of the gauge field with the spin-3/2 field and produces on the
boundary the matrix element of the electromagnetic current of the spin-3/2
field. This matrix element has the same tensorial structure as the one for
the $\Delta$-baryon's electromagnetic current obtained in~\cite{15} within
the QCD framework. From the matching of tensorial structures in matrix element
expressions obtained here and in~\cite{15} we obtain the vertex function
coefficients in integrals over the additional dimension. 
These coefficients allow us to study electromagnetic form factors of $\Delta $ 
baryons in the framework of the hard-wall model that we perform in Section~\ref{sec4}.

\section{Hard-wall model for spin-3/2 field}
\label{sec2}

Let us express the main features of the hard-wall model of the AdS/QCD
described in~\cite{11}. Gravity theory in this model is given by a
5-dimensional (5D) anti de-Sitter space (AdS$_{5}$) and with the metric
chosen in Poincare coordinates as
\begin{equation}
ds^{2}=\frac{1}{z^{2}}\left( \eta _{\mu \nu }dx^{\mu }dx^{\nu
}-dz^{2}\right) .  \label{1}
\end{equation}
Here, $z$ is the fifth coordinate and it extends from $0$ to $\infty $,
which are called, correspondingly, the ultraviolet (UV) and the infrared
(IR) boundaries of the AdS space. $\eta _{\mu \nu }$ is the metric of 4D
flat Minkovski space $\left( \eta _{\mu \nu }=diag\left( 1,-1,-1,-1\right)
,\ \mu =0,1,2,3\right) $. In the hard-wall model $z$ is cut off at the
bottom by $\varepsilon $ $(\varepsilon\rightarrow0)$ and at the top by $
z_{m}=1/\Lambda_{QCD}$, ($\varepsilon \leq z\leq z_{m}$), where $
\Lambda_{QCD}$ corresponds to the confinement scale of QCD. The latter
cutoff breaks conformal symmetry at a slice of the AdS$_{5}$ space.
Hereafter, under the boundary we shall mean the cut off IR boundary.

In this background we introduce a gauge filed $\mathcal{V}^{M}\left(
x,z\right) $ and a spin-3/2 field $\Psi _{A}$ that interact with each other.
The gauge field $\mathcal{V}_{M}^{a}\left( x,z\right) $ in the bulk theory
corresponds to the current of spin-3/2 field $J_{\mu }^{a}\left( x\right) =
\overline{\psi}_{\beta}\left( x\right) \gamma _{\mu }\mathcal{T}
^{a}\psi^{\beta}\left( x\right) $ on the boundary. The flavor symmetry group
for the model is $SU(2)_{L}\times SU(2)_{R}$ and the gauge symmetry group is
$SU(2)$. The gauge field $\mathcal{V}^{M}$ has left $\mathcal{V}_{L}^{M}$
and right $\mathcal{V}_{R}^{M} $ components, transforming under flavor
symmetry as representations of $SU(2)_{L}$ and $SU(2)_{R}$ groups
respectively.

Let us write the bulk-to-boundary propagator for the gauge field, which is
called profile function and the solution to equation of motion for the
Rarita-Schwinger field in the background (\ref{1}).

\subsection{Bulk-to-boundary propagator for gauge field}

The action for the bulk gauge field is written in the following form~\cite
{12}:
\begin{equation}
S=-\frac{1}{2g_{5}^{2}}\int d^{5}x\sqrt{G}Tr\left( \mathcal{F}_{L}^{2}+
\mathcal{F}_{R}^{2}\right) ,  \label{2}
\end{equation}
where $\mathcal{F}_{MN}=\partial _{M}\mathcal{V}_{N}-\partial _{N}\mathcal{V}
_{M}-i\left[ \mathcal{V}_{M},\mathcal{V}_{N}\right] $, $G$ is the
determinant of the metric tensor and $g_{5}=2\pi $. In the axial-like gauge
the fifth component of $\mathcal{V}_{M}$ is gauged away, $\mathcal{V}_{z}=0$
. In the momentum space, the $z$-dependence of $\mathcal{V}_{\mu }\left(
x,z\right) $ is separated by the anzats:
\begin{equation}
\widetilde{\mathcal{V}}_{\mu }\left( q,z\right) =\mathcal{V}_{\mu }\left(
q\right) \frac{V\left( q,z\right) }{V\left( q,\varepsilon \right) }.
\label{3}
\end{equation}
The equation of motion over $z$ leads to the next equation for $V\left(
q,z\right)$,
\begin{equation}
z\partial _{z}\left( \frac{1}{z}\partial _{z}V\left( q,z\right) \right)
+q^{2}V\left( q,z\right) =0  \label{4}
\end{equation}
and solution to this equation is expressed via Bessel functions $J_{m}$ and $
Y_{m}$~\cite{12,10}:\
\begin{equation}
V\left( q,z\right) =\frac{\pi }{2}qz\left[ \frac{Y_{0}\left( qz_{m}\right) }{
J_{0}\left( qz_{m}\right) }J_{1}\left( qz\right) +Y_{1}\left( qz\right) 
\right] .  \label{5}
\end{equation}
Remark, the bulk to boundary propagator obtained here is the propagator for the free gauge field. In the hard-wall model a gauge field interacts with the spin 3/2 field and this should lead to changes in the equation of motion of gauge field. Here we shall neglect this backreaction of spin 3/2 field and use this bulk to boundary propagator in our calculations.

\subsection{Rarita-Schwinger field in the bulk}

As is known from the AdS/CFT correspondence of spin-3/2 fields~\cite{13,11}
that in order to obtain one spin-3/2 field with left- and right-handed
components in the boundary theory one needs to introduce two
Rarita-Schwinger (R-S) fields $\Psi _{1}^{M}$ and $\Psi _{2}^{M}$ in the
bulk theory. One of the bulk R-S fields gives left-handed component of the
boundary field and the second one gives the right-handed component of this
field. On the boundary extra components of bulk fields are eliminated by
boundary conditions, that give a mass spectrum of excited states of this
field. For a clearness let us present in brief some formulas from this
AdS/CFT correspondence~\cite{11}.

The action for the Rarita-Schwinger field $\Psi _{A}$ in AdS$_{5}$ is the
extension of the 4D-action into 5D and is given as
\begin{equation}
S=\int d^{5}x\sqrt{G}\left( i\overline{\Psi }_{A}\Gamma ^{ABC}D_{B}\Psi
_{C}-m_{1}\overline{\Psi }_{A}\Psi ^{A}-m_{2}\overline{\Psi }_{A}\Gamma
^{AB}\Psi _{B}\right) .  \label{6}
\end{equation}
Here, $\Gamma ^{ABC}$ and $\Gamma ^{AB}$ are antisymmetric products of $
\Gamma $ matrices: $\Gamma ^{ABC}=\frac{1}{3}\sum_{perm}\left( -1\right)
^{p}\Gamma ^{A}\Gamma ^{B}\Gamma ^{C}$ $=\frac{1}{2}\left( \Gamma ^{B}\Gamma
^{C}\Gamma ^{A}-\Gamma ^{A}\Gamma ^{C}\Gamma ^{B}\right)$, $\Gamma ^{AB}=
\frac{1}{2}\left( \Gamma ^{A}\Gamma ^{B}-\Gamma ^{B}\Gamma ^{A}\right)$. The
covariant derivative $D_{B}$ is defined by the following shift:
\begin{equation*}
D_{B}=\partial _{B}-\frac{i}{4}\omega _{B}^{MN}\Sigma _{MN}-i\left( \mathcal{
\ V}_{L}^{a}\right) _{B}\mathcal{T}^{a},
\end{equation*}
where $\Sigma _{MN}=$ $\frac{1}{i}\Gamma _{MN}$. For the metric (\ref{1})
the vielbein $e_{M}^{A}$ has chosen as $e_{M}^{A}=\frac{1}{z}\eta _{M}^{A}$
and the spin connection $\omega _{B}^{MN}$ has following non-zero components
$\omega _{\mu }^{5A}=-\omega _{\mu }^{A5}=\frac{1}{z}\delta _{\mu }^{A}$ $
\left( \mu =0,1,2,3\right) $. When transforming from the non-coordinate
frame into the coordinate one (orthogonal frame) the vector-spinor $\Psi
_{A} $ transforms via inverse vielbein $e_{A}^{M}=z\eta _{A}^{M}$ as $\Psi
_{A}=e_{A}^{M}\Psi _{M}$.

The equation of motion obtained from the action (\ref{6}) will give us 5D
Rarita-Schwinger equations in the AdS$_{5}$ space (\ref{1}):
\begin{equation}
i\Gamma ^{A}\left( D_{A}\Psi _{B}-D_{B}\Psi _{A}\right) -m_{-}\Psi _{B}+
\frac{m_{+}}{3}\Gamma _{B}\Gamma ^{A}\Psi _{A}=0,  \label{7}
\end{equation}
where $m_{\pm }=m_{1}\pm m_{2}$ are masses of spinor harmonics or the R-S
field on $S^{5}$ of $AdS_{5}\times S^{5}$~\cite{14}. For $\Gamma $ matrices
it is convenient to use chirality basis~\cite{11},
\begin{equation}
\Gamma ^{5}=-i\gamma ^{5}=\left(
\begin{array}{cc}
-i & 0 \\
0 & i
\end{array}
\right) ,\quad \Gamma ^{0}=\left(
\begin{array}{cc}
0 & -1 \\
-1 & 0
\end{array}
\right) ,\quad \Gamma ^{i}=\left(
\begin{array}{cc}
0 & \sigma ^{i} \\
-\sigma ^{i} & 0
\end{array}
\right) ,\quad \left( i=1,2,3\right) .\quad  \label{8}
\end{equation}

In the four-dimensional theory, the R-S field $\psi _{\mu }$ is a 4D vector
-spinor and contains extra components, which correspond to the states of
spin 1/2 particle. These extra components are projected out by application
of the additional Lorentz-covariant constraint
\begin{equation*}
\gamma ^{\mu }\psi _{\mu }=0.
\end{equation*}
In a five-dimensional theory, it is necessary to impose similar condition in
order to project out extra components of the 5D R-S filed $\Psi _{M}$
corresponding to the states of the spin-1/2 field
\begin{equation}
e_{A}^{M}\Gamma ^{A}\Psi _{M}=0.  \label{9}
\end{equation}
This constraint, together with the free equation of motion, gives another
constraint $\partial ^{M}\Psi _{M}=0$. The 5D Rarita-Schwinger filed $\Psi
_{M}$ has one more extra spin-1/2 component $\Psi _{z}$. It can be gauged
away in the unitary gauge or it can be mapped to the longitudinal component
of the massive spin-1 vector mesons in the boundary theory. In this model,
there is no extra spinor states in the boundary theory into which the $\Psi
_{z}$ could be mapped. In ref.~\cite{11}, the condition $\Psi _{z}=0$ was
chosen in order to eliminate this extra spin-1/2 degree of freedom and we
shall follow it. Taking into account these conditions in Rarita-Schwinger
equation (\ref{7}) gives the next set of Dirac equations for the remaining
components
\begin{equation}
\left( iz\Gamma ^{A}\partial _{A}+2i\Gamma ^{5}-m_{-}\right) \Psi _{\mu }=0.
\label{10}
\end{equation}
It is useful to make further calculations in left and right components of
vector-spinor which have properties $\gamma ^{5}\Psi _{L}^{\mu }=\Psi
_{L}^{\mu }$ and $\gamma ^{5}\Psi _{R}^{\mu }=-\Psi _{R}^{\mu }$. Fourier
transformation for these bulk vector-spinors is written as following
\begin{equation}
\Psi _{L,R}^{\mu }\left( x,z\right) =\int d^{4}p\ e^{-ip\cdot
x}F_{L,R}\left( p,z\right) \psi _{L,R}^{\mu }\left( p\right)  \label{11}
\end{equation}
and the 4D vector-spinor $\psi ^{\mu }\left( p\right) $ obeys 4D Dirac
equation
\begin{equation}
\not{\!}{p}\psi ^{\mu }\left( p\right) =\left\vert p\right\vert \psi ^{\mu
}\left( p\right) .  \label{12}
\end{equation}
Here, $\left\vert p\right\vert =\sqrt{p^{2}}$ for a time-like four-momentum $
p $. Then the 5D Dirac equation (\ref{11}) will lead to equations for $
F_{L,R}$ over the fifth coordinate $z$:
\begin{equation}
\left( \partial _{z}^{2}-\frac{4}{z}\partial _{z}+\frac{6\pm m_{-}-m_{-}^{2}
}{z^{2}}\right) F_{L,R}=-p^{2}F_{L,R}.  \label{13}
\end{equation}
The normalizable solution to this equation for non-zero modes $(|p|\neq 0)$
is expressed in terms of Bessel functions of first kind
\begin{equation}
F_{L,R}=C_{L,R}z^{5/2}J_{m_{-}\mp \frac{1}{2}}\left( \left\vert p\right\vert
z\right) ,  \label{14}
\end{equation}
where $C_{L,R}$ are normalization constants. Value of $m_{-}$ can be found
from the relation $\left\vert m_{-}\right\vert =\Delta _{3/2}-2$, which is
given by the AdS/CFT correspondence of the R-S field. Scaling dimension $
\Delta _{3/2}$ for the spin-3/2 composite operator is $\Delta _{3/2}=9/2$~
\cite{13} and $\left\vert m_{-}\right\vert =5/2$. For the left and the right
R-S fields the mass $\left\vert m_{-}\right\vert $ has values $m_{-}=\pm 5/2$
, correspondingly.

In order to get only left-handed component of the R-S field on the boundary
of the AdS space, we should impose a boundary condition on $\Psi ^{M}$ at $
z=z_{m} $, which eliminates the right-handed component of this vector-spinor
on that boundary:
\begin{equation}
\Psi _{R}^{M}\left( x,z_{m}\right) =0.  \label{15}
\end{equation}
This condition gives a mass spectrum $m_{n}$ for the boundary spin-3/2
field, which is expressed in terms of zeros of the Bessel function $
J_{m_{-}+ \frac{1}{2}}\left( \left\vert p\right\vert z\right)$. Then
boundary condition (\ref{15}) will lead to
\begin{equation*}
J_{3}\left( M_{n}z_{m}\right) =0
\end{equation*}
and the spectrum of excited states will be
\begin{equation}
M_{n}=\frac{\alpha _{n}^{(3)}}{z_{m}}.  \label{16}
\end{equation}
Here $\alpha _{n}^{(3)}$ are zeros of the Bessel function $J_{3}\left(
x\right)$.

In order to get the right-handed component on the boundary we introduce
another spin-3/2 field in the bulk of the AdS space-time. Obviously, all
formulas for the left-handed component are applied to the second R-S field.
This time the left-handed component of the bulk field $\Psi ^{M}$ is
eliminated by the boundary condition:
\begin{equation}
\Psi _{L}^{M}\left( x,z_{m}\right) =0.  \label{17}
\end{equation}
Remark, for this component it should be made replacement $m_{-}\rightarrow
-m_{-}$ in the formulas above. The condition (\ref{17}) leads to
\begin{equation*}
J_{-3}\left( M_{n}z_{m}\right) =0
\end{equation*}
and this gives the same spectrum of excited states for the left-handed
components of the field $\Psi ^{M}$ as for the right-handed components. The
normalization constants $C_{L,R}$ were found in~\cite{8} and are equal to
\begin{equation*}
|C_{L,R}|=\frac{\sqrt{2}}{z_{m}J_{2}\left( M_{n}z_{m}\right) }.
\end{equation*}

\section{Electromagnetic current matrix element for $\Delta$
baryons}
\label{sec3}

\subsection{$\Delta $ baryons and electromagnetic form factors}

The model which was described above is applicable to any spin-3/2 fields
that differ from one another by some other quantum number. 
The simplest known  
realistic model for spin-3/2 particles is the model gased on the gauge group $
SU(2)$ and the flavor group $SU(2)_{L}\times SU(2)_{R}$. These are states of
the spin-3/2 field with isospin-3/2, \textit{i.e.}, the multiplet of $\Delta$
baryons. There are four kinds of $\Delta $ baryons ( $\Delta ^{++},$ $\Delta
^{+},$ $\Delta ^{0}$, $\Delta ^{-}$ ) which differ from each other by the
isospin projection. In order to get a difference in the current matrix
element for different kinds of those baryons, we should take into account
the known isotopic structure of the electromagnetic current of $\Delta $
baryons. To this end we can utilize formulas for the isospin operator and
wave functions for $\Delta $ baryons used in~\cite{16,17}. The
electromagnetic current of $\Delta $ baryons can be divided into isoscalar
and isovector parts

\begin{equation}
J_{\mu }=J_{\mu }^{(s)}\frac{I}{2}+J_{\mu }^{(v)}\frac{\mathcal{T}^{(3)}}{2}
\label{18}
\end{equation}
with
\begin{equation}
\quad \mathcal{T}^{(3)}=\left(
\begin{array}{cccc}
3 & 0 & 0 & 0 \\
0 & 1 & 0 & 0 \\
0 & 0 & -1 & 0 \\
0 & 0 & 0 & -3
\end{array}
\right) .  \label{19}
\end{equation}
Here, $I$ is the four dimensional unit matrix and $\frac{1}{2}\mathcal{T}
^{(3)}$ is the isospin operator. The $\mathcal{T}^{(3)}$ and other basic
matrices $\mathcal{T}^{(1),(2)}$ can transit to Pauli matrices by means of
spin-3/2 to 1/2 transition matrices, explicit form of which can be found in
\cite{17}. In this representation, which is called isoquartet
representation, the isotopic part $\zeta ^{\left( i\right) }$ of the wave
function of $\Delta $ baryons are eigenvectors of $\mathcal{T}^{(3)}$
operator:
\begin{equation}
\zeta ^{\left( 1\right) }=\left(
\begin{array}{c}
1 \\
0 \\
0 \\
0
\end{array}
\right) ,\ \zeta ^{\left( 2\right) }=\left(
\begin{array}{c}
0 \\
1 \\
0 \\
0
\end{array}
\right) ,\ \zeta ^{\left( 3\right) }=\left(
\begin{array}{c}
0 \\
0 \\
1 \\
0
\end{array}
\right) ,\ \zeta ^{\left( 4\right) }=\left(
\begin{array}{c}
0 \\
0 \\
0 \\
1
\end{array}
\right) .  \label{20}
\end{equation}
Eigenvalues of corresponding eigenstates are $\mathcal{T}_{i}^{(3)}=3,\ 1,\
-1,\ -3$. Electric charges $Q_{i}$ of these baryons are defined by
eigenvalues of the charge operator
\begin{equation}
Q=\frac{I}{2}+\frac{1}{2}\mathcal{T}^{(3)}.  \label{21}
\end{equation}
$\Delta ^{++},$ $\Delta ^{+},$ $\Delta ^{0}$ and $\Delta ^{-}$ baryons have
a charge $Q_{1}=2$, $Q_{2}=1$, $Q_{3}=0$, and $Q_{4}=-1$ respectively.
Magnetic moments of these baryons also have isoscalar and isovector parts:
\begin{equation*}
\mu ^{(i)}=\mu _{s}\frac{1}{2}+\mu _{v}\frac{\mathcal{T}_{i}^{(3)}}{2}
\end{equation*}
There are two following relations between $\mu ^{(i)}$:
\begin{eqnarray}
\mu ^{(1)}+\mu ^{(3)} &=&2\mu ^{(2)}  \label{22} \\
\mu ^{(4)}+\mu ^{(2)} &=&2\mu ^{(3)}  \notag
\end{eqnarray}

In the field theory framework the electromagnetic form factors of spin-3/2
particles had been studied in~\cite{15}. The general structure of the matrix
element of the electromagnetic current of these particles was found to be of
the following form:
\begin{equation}
\left\langle p\prime ,s\prime \right\vert j^{\mu }\left( 0\right) \left\vert
p,s\right\rangle =\overline{u}_{\alpha }\left( p\prime ,s\prime \right)
O^{\alpha \mu \beta }u_{\beta }\left( p,s\right) ,  \label{23}
\end{equation}
where $p\prime ,p,s\prime ,s$ are momenta and spin states of a final and an
initial baryon, $u_{\alpha }$ is a vector-spinor describing it. The
Lorentz-covariant tensor $O^{\alpha \mu \beta }$ has explicit form:
\begin{equation}
O^{\alpha \mu \beta }=g^{\alpha \beta }\left[ a_{1}\gamma ^{\mu }+\frac{
a_{2} }{2m}P^{\mu }\right] +\frac{q^{\alpha }q^{\beta }}{\left( 2m\right)
^{2}} \left[ c_{1}\gamma ^{\mu }+\frac{c_{2}}{2m}P^{\mu }\right] ,
\label{24}
\end{equation}
where $P=p\prime +p$ and $q=p\prime -p.$ Coefficients $a_{i}$ and $c_{i}$
are independent covariant vertex function coefficients. Multipole
form-factors of the $\Delta $ baryons were expressed by means of these
coefficients in the following way:
\begin{equation}
G_{E0}\left( q^{2}\right) =\left( 1+\frac{2}{3}\tau \right) \left[
a_{1}+\left( 1+\tau \right) a_{2}\right] -\frac{1}{3}\tau \left( 1+\tau
\right) \left[ c_{1}+\left( 1+\tau \right) c_{2}\right] \, ,  \label{25 a}
\end{equation}
\begin{equation}
G_{E2}\left( q^{2}\right) =\left[ a_{1}+\left( 1+\tau \right) a_{2}\right] -
\frac{1}{2}\tau \left( 1+\tau \right) \left[ c_{1}+\left( 1+\tau \right)
c_{2}\right] \, ,  \label{25 b}
\end{equation}
\begin{equation}
G_{M1}\left( q^{2}\right) =\left( 1+\frac{4}{5}\tau \right) a_{1}-\frac{2}{5}
\tau \left( 1+\tau \right) c_{1} \, ,  \label{25 c}
\end{equation}
\begin{equation}
G_{M3}\left(q^{2}\right) =a_{1}-\frac{1}{2}\left( 1+\tau \right) c_{1}\, ,
\label{25 d}
\end{equation}
where $\tau =-q^{2}/\left( 2m\right) ^{2}$. $G_{E0},$ $G_{E2},$ $G_{M1},$
and $G_{M3}$ are called charge $\left( E0\right) $, electroquadrupole $
\left( E2\right) $, magnetic-dipole $\left( M1\right) $ and
magnetic-octupole $\left( M3\right) $ form factors respectively. Thus, it is
enough to find $a_{i}$ and $c_{i}$ coefficients in order to study form
factors (\ref{25 a})-( \ref{25 d}) and related physical quantities. Our aim
is to find within hard-wall model of AdS/QCD the expressions corresponding
to these coefficients.

\subsection{ Vertex function coefficients in hard-wall model}

In order to obtain an expression for the electromagnetic interaction vertex
of the $\Delta $ baryons within AdS/QCD model, we should start from the
interaction Lagrangian of the Rarita-Schwinger field with the gauge field in
the bulk theory. Note that this Lagrangian besides simple interaction term
will contain different interaction terms which should lead to contributions
of interactions with different multipole moments in the boundary theory. So,
it is awaited that such Lagrangian will be significantly complicated. Let us
determine the interaction Lagrangian in the AdS$_{5}$ space-time with metric
(\ref{1}). First, it should match the gauge symmetry group in the bulk
theory with the flavor symmetry group in the boundary theory. According to
the AdS/CFT correspondence principle, the gauge symmetry of the vector field
in the bulk theory is the same as the flavor symmetry of the corresponding
current in the boundary theory. So, the isotopic structure of the vector
field $\mathcal{V}_{N}$ in the bulk of the AdS$_{5}$ should be same as the
isotopic structure of the barionic current on the boundary of the AdS$_{5}$,
\textit{i.e.}, be as following
\begin{equation}
\mathcal{V}_{N}=V_{N}^{(s)}\frac{I}{2}+V_{N}^{(v)}\frac{\mathcal{T}^{(3)}}{2}
.  \label{26}
\end{equation}
Here $V_{N}^{(s)}$ and $V_{N}^{(v)}$ are isoscalar and isovector parts of
the vector field. In the simplest case (for the photon) these parts are
equal $V_{N}^{(s)}=V_{N}^{(v)}=V_{N}$ and the isotopic part of $\mathcal{V}
_{N}$ is factored out as the charge operator $Q$:
\begin{equation*}
\mathcal{V}_{N}=V_{N}Q.
\end{equation*}
In this representation the field strength tensor $\mathcal{F}_{MN}$ becomes
the one for abelian field $\mathcal{F}_{MN}=\partial _{M}\mathcal{V}
_{N}-\partial _{N}\mathcal{V}_{M}$ and has the same isotopic structure as $
\mathcal{V}_{N}:$
\begin{equation}
\mathcal{F}_{MN}=F_{MN}\frac{I}{2}+F_{MN}\frac{\mathcal{T}^{(3)}}{2}=F_{MN}Q
\label{27}
\end{equation}
Note that the isotopic part of the bulk Rarita-Schwinger field is defined in
the same representation as in the wave function of $\Delta $ baryons,
\textit{i.e.}, we define this part in the isoquartet representation:
\begin{equation}
\Psi ^{M}=\psi ^{M}\zeta ^{\left( i\right) }.  \label{28}
\end{equation}
It is obvious that in any terms of interaction including these fields, like $
\mathcal{V}_{N}\Psi ^{M}$, $\mathcal{F}_{MN}\Psi ^{M}$ and so on, the value
of charge $Q_{i}$ will be factored out:
\begin{equation*}
\mathcal{V}_{N}\Psi ^{M}=Q_{i}V_{N}\psi ^{M}\zeta ^{\left( i\right) }.
\end{equation*}

Interaction Lagrangian in the bulk is constructed by composing hermitian
Lorentz scalars using the 5D R-S fields $\Psi _{1,2}^{M}$, $\overline{\Psi }
_{1,2}^{M}$, the gauge field $\mathcal{V}_{N}$, the field stress tensor $
\mathcal{F}_{MN}$ and its dual tensor $\widetilde{\mathcal{F}}^{MBN}$, their
derivatives $\partial _{M}\Psi _{1,2}^{A}$, $\partial _{M} \overline{\Psi }
_{1,2}^{A}$, $\partial ^{A}\mathcal{F}^{MB}$, $\partial ^{A} \widetilde{
\mathcal{F}}^{MBN}$ and matrices $\Gamma ^{A}$, $\Gamma ^{AB}$ and $\Gamma
^{ABC}$. Main requirement for constructing a Lagrangian term is to obtain
from it on the boundary a matrix element, which has tensorial structure
coinciding with one of structures in (\ref{24}). Let us collect all possible
Lorentz scalars, which obey this requirement. It will consist of the
following terms of interaction of the vector-spinor with the gauge field:

a) terms describing simple interaction vertices
\begin{equation}
\mathcal{L}_{0}=\overline{\Psi }_{1}^{M}\Gamma ^{N}\mathcal{V}_{N}\Psi
_{1M}+ \overline{\Psi }_{2}^{M}\Gamma ^{N}\mathcal{V}_{N}\Psi _{2M},
\label{29}
\end{equation}

b) terms describing an interaction with magnetic dipole moment
\begin{equation*}
\mathcal{L}_{1}=\left\{ \eta ^{(s)}\left[ \overline{\Psi }_{1}^{M}\Gamma
^{NP}F_{NP}\frac{I}{2}\Psi _{1M}-\overline{\Psi }_{2}^{M}\Gamma ^{NP}F_{NP}
\frac{I}{2}\Psi _{2M}\right] \right.
\end{equation*}
\begin{equation}
\left. +\eta ^{(v)}\left[ \overline{\Psi }_{1}^{M}\Gamma ^{NP}F_{NP}\frac{
\mathcal{T}^{(3)}}{2}\Psi _{1M}-\overline{\Psi }_{2}^{M}\Gamma ^{NP}F_{NP}
\frac{\mathcal{T}^{(3)}}{2}\Psi _{2M}\right] \right\} ,  \label{30}
\end{equation}
Coefficients $\eta ^{(s)}$ and $\eta ^{(v)}$ introduced here take into
account difference in contributions of isoscalar and isovector parts of the
magnetic moment of the $\Delta $ baryon~\cite{9}.

Other Lorentz invariants are constructed by means of fields and derivatives,
which produce one (or more) of required tensorial structures on the
boundary. We classify them as following:

c) terms constructed by $\mathcal{V}^{M}$
\begin{equation}
\mathcal{L}_{2}=ik_{2}\left[ \left( \partial _{M}\overline{\Psi }
_{1A}\right) \mathcal{V}^{M}\Psi _{1}^{A}-\overline{\Psi }_{1A}\mathcal{V}
^{M}\left( \partial _{M}\Psi _{1}^{A}\right) -\left( 1\rightarrow 2\right) 
\right] ,  \label{31}
\end{equation}
\begin{equation}
\mathcal{L}_{3}=ik_{3}\left[ \left( \partial _{M}\overline{\Psi }
_{1A}\right) \Gamma ^{AB}\mathcal{V}^{M}\Psi _{1B}-\overline{\Psi }
_{1A}\Gamma ^{AB}\mathcal{V}^{M}\left( \partial _{M}\Psi _{1B}\right)
-\left( 1\rightarrow 2\right) \right] ,  \label{32}
\end{equation}
\begin{equation}
\mathcal{L}_{4}=k_{4}\left[ \overline{\Psi }_{1A}\Gamma ^{AMN}\mathcal{V}
_{M}\Psi _{1N}-\left( 1\rightarrow 2\right) \right] ,  \label{33}
\end{equation}

d) terms constructed by ${\mathcal{F}}^{MN}$:
\begin{equation}
\mathcal{L}_{5}=\frac{ik_{5}}{2}\left[ \overline{\Psi }_{1M}{\mathcal{F}}
^{MN}\Psi _{1N}-\overline{\Psi }_{2M}{\mathcal{F}}^{MN}\Psi _{2N}\right]
+h.c.,  \label{34}
\end{equation}
\begin{equation}
\mathcal{L}_{6}=\frac{ik_{6}}{2}\left[ \left( \partial _{M}\overline{\Psi }
_{1A}\right) \mathcal{F}^{MN}\left( \partial _{N}\Psi _{1}^{A}\right)
-\left( \partial _{N}\overline{\Psi }_{1A}\right) \mathcal{F}^{MN}\left(
\partial _{M}\Psi _{1}^{A}\right) -\left( 1\rightarrow 2\right) \right]
+h.c.,  \label{35}
\end{equation}
\begin{equation}
\mathcal{L}_{7}=\frac{ik_{7}}{2}\left[ \left( \partial _{M}\overline{\Psi }
_{1A}\right) \mathcal{F}^{MN}\left( \partial ^{A}\Psi _{1N}\right) -\left(
\partial ^{A}\overline{\Psi }_{1A}\right) \mathcal{F}^{MN}\left( \partial
_{M}\Psi _{1N}\right) -\left( 1\rightarrow 2\right) \right] +h.c.,
\label{36}
\end{equation}
\begin{equation}
\mathcal{L}_{8}=\frac{ik_{8}}{2}\left[ \left( \partial _{A}\overline{\Psi }
_{1M}\right) \Gamma ^{A}\mathcal{F}^{MN}\Psi _{1N}-\overline{\Psi }
_{1M}\Gamma ^{A}\mathcal{F}^{MN}\left( \partial _{A}\Psi _{1N}\right)
-\left( 1\rightarrow 2\right) \right] +h.c.,  \label{37}
\end{equation}
\begin{equation}
\mathcal{L}_{9}=\frac{k_{9}}{2}\left[ \left( \partial _{A}\overline{\Psi }
_{1B}\right) \Gamma ^{AMN}\mathcal{F}_{MN}\Psi _{1}^{B}-\overline{\Psi }
_{1B}\Gamma ^{AMN}\mathcal{F}_{MN}\left( \partial _{A}\Psi _{1}^{B}\right)
+\left( 1\rightarrow 2\right) \right] +h.c.,  \label{38}
\end{equation}
e) terms constructed by $\widetilde{\mathcal{F}}^{AMN}$
\begin{equation}
\mathcal{L}_{10}=\frac{ik_{10}}{2}\left[ \overline{\Psi }_{1A}\Gamma _{M}
\widetilde{\mathcal{F}}^{AMN}\Psi _{1N}-\left( 1\rightarrow 2\right) \right]
+h.c.,  \label{39}
\end{equation}
\begin{equation}
\mathcal{L}_{11}=\frac{k_{11}}{2}\left[ \left( \partial ^{C}\overline{\Psi }
_{1A}\right) \Gamma _{CM}\widetilde{\mathcal{F}}^{AMN}\Psi _{1N}-\overline{
\Psi }_{1A}\Gamma _{CM}\widetilde{\mathcal{F}}^{AMN}\left( \partial ^{C}\Psi
_{1N}\right) -\left( 1\rightarrow 2\right) \right] +h.c.,  \label{40}
\end{equation}
\begin{equation}
\mathcal{L}_{12}=\frac{ik_{12}}{2}\left[ \overline{\Psi }_{1}^{B}\Gamma
_{BAM}\widetilde{\mathcal{F}}^{AMN}\Psi _{1N}-\left( 1\rightarrow 2\right) 
\right] +h.c.,  \label{41}
\end{equation}
\begin{equation}
\mathcal{L}_{13}=\frac{ik_{13}}{2}\left[ \overline{\Psi }_{1}^{B}\Gamma
_{AMN}\widetilde{\mathcal{F}}^{AMN}\Psi _{1B}-\left( 1\rightarrow 2\right) 
\right] +h.c.,  \label{42}
\end{equation}
f) terms constructed by $\partial ^{A}\mathcal{F}^{MB}$
\begin{equation}
\mathcal{L}_{14}=\frac{ik_{14}}{2}\left[ \left( \partial _{M}\overline{\Psi }
_{1A}\right) \left( \partial ^{A}\mathcal{F}^{MB}\right) \Psi _{1B}-
\overline{\Psi }_{1A}\left( \partial ^{A}\mathcal{F}^{MB}\right) \left(
\partial _{M}\Psi _{1B}\right) -\left( 1\rightarrow 2\right) \right] +h.c.,
\label{43}
\end{equation}
\begin{equation}
\mathcal{L}_{15}=\frac{k_{15}}{2}\left[ \overline{\Psi }_{1A}\Gamma
_{M}\left( \partial ^{A}\mathcal{F}^{MB}\right) \Psi _{1B}+\left(
1\rightarrow 2\right) \right] +h.c.,  \label{44}
\end{equation}
g) terms constructed by $\partial ^{A}\widetilde{\mathcal{F}}^{MBN}$
\begin{equation}
\mathcal{L}_{16}=\frac{ik_{16}}{2}\left[ \left( \partial _{B}\overline{\Psi }
_{1A}\right) \Gamma _{M}\left( \partial ^{A}\widetilde{\mathcal{F}}
^{MBN}\right) \Psi _{1N}+\overline{\Psi }_{1A}\Gamma _{M}\left( \partial ^{A}
\widetilde{\mathcal{F}}^{MBN}\right) \left( \partial _{B}\Psi _{1N}\right)
-\left( 1\rightarrow 2\right) \right] +h.c.  \label{45}
\end{equation}
\begin{equation}
\mathcal{L}_{17}=\frac{k_{17}}{2}\left[ \overline{\Psi }_{1A}\Gamma
_{MB}\left( \partial ^{A}\widetilde{\mathcal{F}}^{MBN}\right) \Psi
_{1N}+\left( 1\rightarrow 2\right) \right] +h.c.,  \label{46}
\end{equation}
\begin{equation}
\mathcal{L}_{18}=\frac{ik_{18}}{2}\left[ \left( \partial _{M}\overline{\Psi }
_{1}^{C}\right) \Gamma _{CAB}\left( \partial ^{A}\widetilde{\mathcal{F}}
^{MBN}\right) \Psi _{1N}+\overline{\Psi }_{1}^{C}\Gamma _{CAB}\left(
\partial ^{A}\widetilde{\mathcal{F}}^{MBN}\right) \left( \partial _{M}\Psi
_{1N}\right) -\left( 1\rightarrow 2\right) \right] +h.c.  \label{47}
\end{equation}
Here, $k_{s}$ are arbitrary real coefficients. Let us remark that because of
isotopic symmetry each Lagrangian term consist of four terms corresponding
to different values of $\mathcal{T}_{i}^{(3)}$ $\left( \mathcal{L}
_{s}=\sum\limits_{i=1}^{4}\mathcal{L}_{s}^{\left( i\right) }\right) $.
The interaction vertex of another multiplet of spin-3/2 fields with the
photon also can be described by these Lagrangian terms after changes of
the isotopic symmetry group. Some terms of chiral symmetry breaking, like following one
(see~\cite{11})
\begin{equation*}
\mathcal{L}_{\chi SB}=\frac{i}{2}k_{1}\left[ \overline{\Psi }
_{1}^{M}X^{3}\Gamma ^{NP}\mathcal{F}_{NP}\Psi _{2M}+\overline{\Psi }
_{2}^{M}\left( X^{\dagger }\right) ^{3}\Gamma ^{NP}\mathcal{F}_{NP}\Psi _{1M}
\right] ,
\end{equation*}
are possible as well. We did not include here those terms into Lagrangian,
since their contributions to the current matrix element are too small in
comparison with other terms due to the $X^{3}$ factor. Thus, the total
interaction Lagrangian of the R-S field with the bulk gauge field will
consist of the sum of all Lagrangian terms $\mathcal{L}_{0}$-$\mathcal{L}
_{18}$:
\begin{equation}
\mathcal{L}_{int}=\sum\limits_{s=0}^{18}\mathcal{L}_{s}  \label{48}
\end{equation}
$S$-matrix element $S_{fi}$ of this interaction, in the first approximation,
is obtained from this Lagrangian by performing the 5D integration:
\begin{equation}
S_{fi}=\int\limits_{\varepsilon }^{z_{m}}dz\sqrt{G}\int d^{4}x\mathcal{L}
_{int}.  \label{49}
\end{equation}
Integration of each Lagrangian term $\mathcal{L}_{s}$ over $x$ and using the
equation of motion (\ref{12}) and the constraint (\ref{9}) leads to momentum
integrals of the product of vector-spinors $\overline{u}_{\alpha }\left(
p\prime \right) $, $u_{\beta }\left( p\right) $ of final and initial states
with a tensor $\tilde{O}_{s}^{\alpha \mu \beta }$. So, $S_{fi}$ in momentum
space is reduced to the following form:
\begin{equation}
S_{fi}=\int d^{4}p^{\prime }d^{4}p\,\overline{u}_{\alpha }\left( p\prime
\right) \varepsilon _{\mu }\tilde{O}^{\alpha \mu \beta }u_{\beta }\left(
p\right) ,  \label{50}
\end{equation}
where $\tilde{O}^{\alpha \mu \beta }$ is the sum of all contributions $
\tilde{O}_{s}^{\alpha \mu \beta }$:
\begin{equation}
\tilde{O}^{\alpha \mu \beta }=\sum\limits_{s=0}^{18}\tilde{O}_{s}^{\alpha
\mu \beta }.  \label{51}
\end{equation}
Here $p\prime $ and $p$ are momenta of final and initial states of R-S
filed, which are states after and before interaction with the gauge field.

For a transparency it is useful to present $\tilde{O}_{s}^{\alpha \mu \beta
} $ separately. The obtained expressions of $\tilde{O}_{s}^{\alpha \mu \beta
}$ are the following ones:
\begin{equation*}
\tilde{O}_{0}^{\alpha \mu \beta }=g^{\alpha \beta }\gamma ^{\mu }\frac{1}{
\sqrt{2}}Q_{i}\int\limits_{0}^{z_{m}}dz\frac{V\left( qz\right) }{z^{2}}\left[
F_{1L}^{\ast }\left( p\prime z\right) F_{1L}\left( pz\right) +F_{1R}^{\ast
}\left( p\prime z\right) F_{1R}\left( pz\right) \right.
\end{equation*}
\begin{equation*}
\left. +F_{2L}^{\ast }\left( p\prime z\right) F_{2L}\left( pz\right)
+F_{2R}^{\ast }\left( p\prime z\right) F_{2R}\left( pz\right) \right]
\end{equation*}
\begin{equation*}
\tilde{O}_{1}^{\alpha \mu \beta }=g^{\alpha \beta }\left[ m\gamma ^{\mu }-
\frac{1}{2}P^{\mu }\right] \frac{1}{\sqrt{2}}\left( \eta ^{(s)}+\eta ^{(v)}
\mathcal{T}_{i}^{(3)}\right) \int\limits_{0}^{z_{m}}dz\frac{V\left(
qz\right) }{z}\times
\end{equation*}
\begin{equation*}
\left[ F_{1L}^{\ast }\left( p\prime z\right) F_{1R}\left( pz\right)
+F_{1R}^{\ast }\left( p\prime z\right) F_{1L}\left( pz\right) -F_{2L}^{\ast
}\left( p\prime z\right) F_{2R}\left( pz\right) -F_{2R}^{\ast }\left(
p\prime z\right) F_{2L}\left( pz\right) \right]
\end{equation*}
\begin{equation*}
+g^{\alpha \beta }\gamma ^{\mu }\frac{1}{\sqrt{2}}\left( \eta ^{(s)}+\eta
^{(v)}\mathcal{T}_{i}^{(3)}\right) \int\limits_{0}^{z_{m}}dz\frac{\partial
_{z}V\left( qz\right) }{z}\times
\end{equation*}
\begin{equation*}
\left[ F_{1L}^{\ast }\left( p\prime z\right) F_{1L}\left( pz\right)
-F_{1R}^{\ast }\left( p\prime z\right) F_{1R}\left( pz\right) -F_{2L}^{\ast
}\left( p\prime z\right) F_{2L}\left( pz\right) +F_{2R}^{\ast }\left(
p\prime z\right) F_{2R}\left( pz\right) \right]
\end{equation*}
\begin{equation*}
\tilde{O}_{2}^{\alpha \mu \beta }=-g^{\alpha \beta }P^{\mu }\frac{1}{2\sqrt{2
}}k_{2}Q_{i}\int\limits_{0}^{z_{m}}dz\frac{V\left( qz\right) }{z^{3}}\times
\end{equation*}
\begin{equation*}
\left[ F_{1L}^{\ast }\left( p\prime z\right) F_{1R}\left( pz\right)
+F_{1R}^{\ast }\left( p\prime z\right) F_{1L}\left( pz\right) -F_{2L}^{\ast
}\left( p\prime z\right) F_{2R}\left( pz\right) -F_{2R}^{\ast }\left(
p\prime z\right) F_{2L}\left( pz\right) \right]
\end{equation*}
\begin{equation*}
\tilde{O}_{3}^{\alpha \mu \beta }=g^{\alpha \beta }P^{\mu }\frac{1}{2\sqrt{2}
}Q_{i}k_{3}\int\limits_{0}^{z_{m}}dz\frac{V\left( qz\right) }{z}\times
\end{equation*}
\begin{equation*}
\left[ F_{1L}^{\ast }\left( p\prime z\right) F_{1R}\left( pz\right)
+F_{1R}^{\ast }\left( p\prime z\right) F_{1L}\left( pz\right) -F_{2L}^{\ast
}\left( p\prime z\right) F_{2R}\left( pz\right) -F_{2R}^{\ast }\left(
p\prime z\right) F_{2L}\left( pz\right) \right]
\end{equation*}
\begin{equation*}
\tilde{O}_{4}^{\alpha \mu \beta }=g^{\alpha \beta }\gamma ^{\mu }\frac{1}{
\sqrt{2}}Q_{i}k_{4}\int\limits_{0}^{z_{m}}dzV\left( qz\right) \times
\end{equation*}
\begin{equation*}
\left[ F_{1L}^{\ast }\left( p\prime z\right) F_{1R}\left( pz\right)
+F_{1R}^{\ast }\left( p\prime z\right) F_{1L}\left( pz\right) -F_{2L}^{\ast
}\left( p\prime z\right) F_{2R}\left( pz\right) -F_{2R}^{\ast }\left(
p\prime z\right) F_{2L}\left( pz\right) \right]
\end{equation*}
\begin{equation*}
\tilde{O}_{5}^{\alpha \mu \beta }=\left\{ g^{\alpha \beta }\left[ m\left( 1-
\frac{q^{2}}{4m^{2}}\right) \gamma ^{\mu }-\frac{1}{2}P^{\mu }\right] +\frac{
1}{2m}q^{\alpha }q^{\beta }\gamma ^{\mu }\right\} \frac{1}{\sqrt{2}}
k_{5}Q_{i}\int\limits_{0}^{z_{m}}dz\frac{V\left( qz\right) }{z^{3}}\times
\end{equation*}
\begin{equation*}
\left[ F_{1L}^{\ast }\left( p\prime z\right) F_{1R}\left( pz\right)
+F_{1R}^{\ast }\left( p\prime z\right) F_{1L}\left( pz\right) -F_{2L}^{\ast
}\left( p\prime z\right) F_{2R}\left( pz\right) -F_{2R}^{\ast }\left(
p\prime z\right) F_{2L}\left( pz\right) \right]
\end{equation*}
\begin{equation*}
\tilde{O}_{6}^{\alpha \mu \beta }=g^{\alpha \beta }P^{\mu }\frac{1}{2\sqrt{2}
}k_{6}Q_{i}\int\limits_{0}^{z_{m}}dz\frac{1}{z^{3}}\left\{ q^{2}V\left(
qz\right) \left[ F_{1L}^{\ast }\left( p\prime z\right) F_{1R}\left(
pz\right) +F_{1R}^{\ast }\left( p\prime z\right) F_{1L}\left( pz\right)
\right. \right.
\end{equation*}
\begin{equation*}
\left. \left. -F_{2L}^{\ast }\left( p\prime z\right) F_{2R}\left( pz\right)
-F_{2R}^{\ast }\left( p\prime z\right) F_{2L}\left( pz\right) \right]
-\partial _{z}V\left( qz\right) \left[ \partial _{z}F_{1L}^{\ast }\left(
p\prime z\right) F_{1R}\left( pz\right) +\right. \right.
\end{equation*}
\begin{equation*}
\left. \left. +\partial _{z}F_{1R}^{\ast }\left( p\prime z\right)
F_{1L}\left( pz\right) +F_{1L}^{\ast }\left( p\prime z\right) \partial
_{z}F_{1R}\left( pz\right) +F_{1R}^{\ast }\left( p\prime z\right) \partial
_{z}F_{1L}\left( pz\right) -\right. \right.
\end{equation*}
\begin{equation*}
\left. \left. -\partial _{z}F_{2L}^{\ast }\left( p\prime z\right)
F_{2R}\left( pz\right) -\partial _{z}F_{2R}^{\ast }\left( p\prime z\right)
F_{2L}\left( pz\right) -F_{2L}^{\ast }\left( p\prime z\right) \partial
_{z}F_{2R}\left( pz\right) -F_{2R}^{\ast }\left( p\prime z\right) \partial
_{z}F_{2L}\left( pz\right) \right] \right\}
\end{equation*}
\begin{equation*}
\tilde{O}_{7}^{\alpha \mu \beta }=-\left\{ 2g^{\alpha \beta }\left[ m\left(
1-\frac{q^{2}}{4m^{2}}\right) \gamma ^{\mu }-\frac{1}{2}P^{\mu }\right] +
\frac{1}{m}q^{\alpha }q^{\beta }\gamma ^{\mu }\right\} \frac{1}{2\sqrt{2}}
k_{7}Q_{i}\int\limits_{0}^{z_{m}}dz\frac{V\left( qz\right) }{z^{2}}\times
\end{equation*}
\begin{equation*}
\left[ F_{1L}^{\ast }\left( p\prime z\right) F_{1R}\left( pz\right)
+F_{1R}^{\ast }\left( p\prime z\right) F_{1L}\left( pz\right) -F_{2L}^{\ast
}\left( p\prime z\right) F_{2R}\left( pz\right) -F_{2R}^{\ast }\left(
p\prime z\right) F_{2L}\left( pz\right) \right]
\end{equation*}
\begin{equation*}
\tilde{O}_{8}^{\alpha \mu \beta }=\left\{ 2mg^{\alpha \beta }\left[ m\left(
1-\frac{q^{2}}{4m^{2}}\right) \gamma ^{\mu }-\frac{1}{2}P^{\mu }\right]
+q^{\alpha }q^{\beta }\gamma ^{\mu }\right\} \frac{1}{\sqrt{2}}
k_{8}Q_{i}\int\limits_{0}^{z_{m}}dz\frac{V\left( qz\right) }{z^{2}}\times
\end{equation*}
\begin{equation*}
\left[ F_{1L}^{\ast }\left( p\prime z\right) F_{1R}\left( pz\right)
+F_{1R}^{\ast }\left( p\prime z\right) F_{1L}\left( pz\right) -F_{2L}^{\ast
}\left( p\prime z\right) F_{2R}\left( pz\right) -F_{2R}^{\ast }\left(
p\prime z\right) F_{2L}\left( pz\right) \right]
\end{equation*}
\begin{equation*}
\tilde{O}_{9}^{\alpha \mu \beta }=2mg^{\alpha \beta }\left[ 2m\left( 1-\frac{
q^{2}}{4m^{2}}\right) \gamma ^{\mu }-P^{\mu }\right] \frac{1}{\sqrt{2}}
k_{9}Q_{i}\int\limits_{0}^{z_{m}}dzV\left( qz\right) \times
\end{equation*}
\begin{equation*}
\left[ F_{1L}^{\ast }\left( p\prime z\right) F_{1L}\left( pz\right)
+F_{1R}^{\ast }\left( p\prime z\right) F_{1R}\left( pz\right) +F_{2L}^{\ast
}\left( p\prime z\right) F_{2L}\left( pz\right) +F_{2R}^{\ast }\left(
p\prime z\right) F_{2R}\left( pz\right) \right]
\end{equation*}
\begin{equation*}
\tilde{O}_{10}^{\alpha \mu \beta }=-g^{\alpha \beta }\gamma ^{\mu }\frac{1}{2
\sqrt{2}}k_{10}Q_{i}\int\limits_{0}^{z_{m}}dz\frac{\partial _{z}V\left(
qz\right) }{z^{2}}\times
\end{equation*}
\begin{equation*}
\left[ F_{1L}^{\ast }\left( p\prime z\right) F_{1L}\left( pz\right)
-F_{1R}^{\ast }\left( p\prime z\right) F_{1R}\left( pz\right) -F_{2L}^{\ast
}\left( p\prime z\right) F_{2L}\left( pz\right) +F_{2R}^{\ast }\left(
p\prime z\right) F_{2R}\left( pz\right) \right]
\end{equation*}
\begin{equation*}
\tilde{O}_{11}^{\alpha \mu \beta }=-\left\{ g^{\alpha \beta }\frac{q^{2}}{2}
\gamma ^{\mu }-q^{\alpha }q^{\beta }\gamma ^{\mu }\right\} \sqrt{2}
k_{11}Q_{i}\int\limits_{0}^{z_{m}}dz\frac{V\left( qz\right) }{z}\times
\end{equation*}
\begin{equation*}
\left[ F_{1L}^{\ast }\left( p\prime z\right) F_{1R}\left( pz\right)
+F_{1R}^{\ast }\left( p\prime z\right) F_{1L}\left( pz\right) -F_{2L}^{\ast
}\left( p\prime z\right) F_{2R}\left( pz\right) -F_{2R}^{\ast }\left(
p\prime z\right) F_{2L}\left( pz\right) \right]
\end{equation*}
\begin{equation*}
\tilde{O}_{12}^{\alpha \mu \beta }=\left\{ g^{\alpha \beta }\left[ \frac{
q^{2}}{2m}\gamma ^{\mu }-P^{\mu }\right] +\frac{1}{m}q^{\alpha }q^{\beta
}\gamma ^{\mu }\right\} 2\sqrt{2}k_{12}Q_{i}\int\limits_{0}^{z_{m}}dz\left\{
V(qz)\times \right.
\end{equation*}
\begin{equation*}
\left. \left[ F_{1L}^{\ast }\left( p\prime z\right) F_{1R}\left( pz\right)
+F_{1R}^{\ast }\left( p\prime z\right) F_{1L}\left( pz\right) -F_{2L}^{\ast
}\left( p\prime z\right) F_{2R}\left( pz\right) -F_{2R}^{\ast }\left(
p\prime z\right) F_{2L}\left( pz\right) \right] \right.
\end{equation*}
\begin{equation*}
\left. -\partial _{z}V\left( qz\right) \left[ F_{1L}^{\ast }\left( p\prime
z\right) F_{1L}\left( pz\right) -F_{1R}^{\ast }\left( p\prime z\right)
F_{1R}\left( pz\right) -F_{2L}^{\ast }\left( p\prime z\right) F_{2L}\left(
pz\right) +F_{2R}^{\ast }\left( p\prime z\right) F_{2R}\left( pz\right) 
\right] \right\}
\end{equation*}
\begin{equation*}
\tilde{O}_{13}^{\alpha \mu \beta }=g^{\alpha \beta }\gamma ^{\mu }3\sqrt{2}
k_{13}Q_{i}\int\limits_{0}^{z_{m}}dzV\left( qz\right) \times
\end{equation*}
\begin{equation*}
\left[ F_{1L}^{\ast }\left( p\prime z\right) F_{1L}\left( pz\right)
-F_{1R}^{\ast }\left( p\prime z\right) F_{1R}\left( pz\right) -F_{2L}^{\ast
}\left( p\prime z\right) F_{2L}\left( pz\right) +F_{2R}^{\ast }\left(
p\prime z\right) F_{2R}\left( pz\right) \right]
\end{equation*}
\begin{equation*}
\tilde{O}_{14}^{\alpha \mu \beta }=-q^{\alpha }q^{\beta }P^{\mu }\frac{1}{2
\sqrt{2}}Q_{i}k_{14}\int\limits_{0}^{z_{m}}dz\,V\left( qz\right) z^{-3}\left[
F_{1L}^{\ast }\left( p\prime z\right) F_{1R}\left( pz\right) +F_{1R}^{\ast
}\left( p\prime z\right) F_{1L}\left( pz\right) \right.
\end{equation*}
\begin{equation*}
\left. -F_{2L}^{\ast }\left( p\prime z\right) F_{2R}\left( pz\right)
-F_{2R}^{\ast }\left( p\prime z\right) F_{2L}\left( pz\right) \right]
\end{equation*}
\begin{equation*}
\tilde{O}_{15}^{\alpha \mu \beta }=q^{\alpha }q^{\beta }\gamma ^{\mu }\frac{1
}{2\sqrt{2}}Q_{i}k_{15}\int\limits_{0}^{z_{m}}dz\,V\left( qz\right) z^{-2}
\left[ F_{1L}^{\ast }\left( p\prime z\right) F_{1L}\left( pz\right)
+F_{1R}^{\ast }\left( p\prime z\right) F_{1R}\left( pz\right) \right.
\end{equation*}
\begin{equation*}
\left. +F_{2L}^{\ast }\left( p\prime z\right) F_{2L}\left( pz\right)
+F_{2R}^{\ast }\left( p\prime z\right) F_{2R}\left( pz\right) \right]
\end{equation*}
\begin{equation*}
\tilde{O}_{16}^{\alpha \mu \beta }=q^{\alpha }q^{\beta }\gamma ^{\mu }\frac{1
}{\sqrt{2}}Q_{i}k_{16}\int\limits_{0}^{z_{m}}dzz^{-2}\left\{ 2V\left(
qz\right) \left[ \partial _{z}F_{1L}^{\ast }\left( p\prime z\right)
F_{1L}\left( pz\right) -\partial _{z}F_{1R}^{\ast }\left( p\prime z\right)
F_{1R}\left( pz\right) \right. \right.
\end{equation*}
\begin{equation*}
\left. +\partial _{z}F_{2L}^{\ast }\left( p\prime z\right) F_{2L}\left(
pz\right) -\partial _{z}F_{2R}^{\ast }\left( p\prime z\right) F_{2R}\left(
pz\right) \right] -
\end{equation*}
\begin{equation*}
\left. -\partial _{z}V\left( qz\right) \left[ F_{1L}^{\ast }\left( p\prime
z\right) F_{1L}\left( pz\right) -F_{1R}^{\ast }\left( p\prime z\right)
F_{1R}\left( pz\right) -F_{2L}^{\ast }\left( p\prime z\right) F_{2L}\left(
pz\right) +F_{2R}^{\ast }\left( p\prime z\right) F_{2R}\left( pz\right) 
\right] \right\} .
\end{equation*}
\begin{equation*}
\tilde{O}_{17}^{\alpha \mu \beta }=-q^{\alpha }q^{\beta }\gamma ^{\mu }\frac{
5}{2\sqrt{2}}Q_{i}k_{17}\int\limits_{0}^{z_{m}}dz\,V\left( qz\right)
z^{-1}\times
\end{equation*}
\begin{equation*}
\left[ F_{1L}^{\ast }\left( p\prime z\right) F_{1L}\left( pz\right)
+F_{1R}^{\ast }\left( p\prime z\right) F_{1R}\left( pz\right) +F_{2L}^{\ast
}\left( p\prime z\right) F_{2L}\left( pz\right) +F_{2R}^{\ast }\left(
p\prime z\right) F_{2R}\left( pz\right) \right]
\end{equation*}
\begin{equation*}
\tilde{O}_{18}^{\alpha \mu \beta }=\left[ -g^{\alpha \beta }\frac{q^{2}}{2m}
\gamma ^{\mu }+\frac{1}{m}q^{\alpha }q^{\beta }\gamma ^{\mu }\right] 2\sqrt{2
}k_{18}Q_{i}\int\limits_{0}^{z_{m}}dz\partial _{z}V(qz)\times
\end{equation*}
\begin{equation*}
\left\{ \left[ F_{1L}^{\ast }\left( p\prime z\right) F_{1R}\left( pz\right)
+F_{1R}^{\ast }\left( p\prime z\right) F_{1L}\left( pz\right) -F_{2L}^{\ast
}\left( p\prime z\right) F_{2R}\left( pz\right) -F_{2R}^{\ast }\left(
p\prime z\right) F_{2L}\left( pz\right) \right] \right.
\end{equation*}
\begin{equation*}
\left. +2m\left[ F_{1L}^{\ast }\left( p\prime z\right) F_{1L}\left(
pz\right) -F_{1R}^{\ast }\left( p\prime z\right) F_{1R}\left( pz\right)
-F_{2L}^{\ast }\left( p\prime z\right) F_{2L}\left( pz\right) +F_{2R}^{\ast
}\left( p\prime z\right) F_{2R}\left( pz\right) \right] \right.
\end{equation*}
\begin{equation*}
\left. -\left[ \partial _{z}F_{1L}^{\ast }\left( p\prime z\right)
F_{1R}\left( pz\right) +\partial _{z}F_{1R}^{\ast }\left( p\prime z\right)
F_{1L}\left( pz\right) +F_{1L}^{\ast }\left( p\prime z\right) \partial
_{z}F_{1R}\left( pz\right) +F_{1R}^{\ast }\left( p\prime z\right) \partial
_{z}F_{1L}\left( pz\right) \right. \right.
\end{equation*}
\begin{equation*}
\left. \left. -\partial _{z}F_{2L}^{\ast }\left( p\prime z\right)
F_{2R}\left( pz\right) -\partial _{z}F_{2R}^{\ast }\left( p\prime z\right)
F_{2L}\left( pz\right) -F_{2L}^{\ast }\left( p\prime z\right) \partial
_{z}F_{2R}\left( pz\right) -F_{2R}^{\ast }\left( p\prime z\right) \partial
_{z}F_{2L}\left( pz\right) \right] \right\}
\end{equation*}
\begin{equation*}
+\frac{1}{2}q^{\alpha }q^{\beta }\gamma ^{\mu }2\sqrt{2}k_{18}Q_{i}\int
\limits_{0}^{z_{m}}dz\partial _{z}V(qz)\left[ F_{1L}^{\ast }\left( p\prime
z\right) F_{1L}\left( pz\right) -F_{1R}^{\ast }\left( p\prime z\right)
F_{1R}\left( pz\right) -\right.
\end{equation*}
\begin{equation*}
\left. -F_{2L}^{\ast }\left( p\prime z\right) F_{2L}\left( pz\right)
+F_{2R}^{\ast }\left( p\prime z\right) F_{2R}\left( pz\right) \right]
-q^{\alpha }q^{\beta }\gamma ^{\mu }2\sqrt{2}k_{18}Q_{i}\int
\limits_{0}^{z_{m}}dzV(qz)\times
\end{equation*}
\begin{equation*}
\times \left[ \partial _{z}F_{1L}^{\ast }\left( p\prime z\right)
F_{1L}\left( pz\right) -\partial _{z}F_{1R}^{\ast }\left( p\prime z\right)
F_{1R}\left( pz\right) +F_{1L}^{\ast }\left( p\prime z\right) \partial
_{z}F_{1L}\left( pz\right) -F_{1R}^{\ast }\left( p\prime z\right) \partial
_{z}F_{1R}\left( pz\right) \right.
\end{equation*}
\begin{equation*}
\left. -\partial _{z}F_{2L}^{\ast }\left( p\prime z\right) F_{2L}\left(
pz\right) +\partial _{z}F_{2R}^{\ast }\left( p\prime z\right) F_{2R}\left(
pz\right) -F_{2L}^{\ast }\left( p\prime z\right) \partial _{z}F_{2L}\left(
pz\right) +F_{2R}^{\ast }\left( p\prime z\right) \partial _{z}F_{2R}\left(
pz\right) \right]
\end{equation*}
Here we have considered initial and final R-S fields on mass shell ( $
\left\vert p\right\vert =m=\left\vert p^{\prime }\right\vert $ ), since $
\Delta $ baryons with which we want to match R-S fields are on the mass
shell as well. In calculations of some terms we used the formulas (4a) and
(4b) from~\cite{15} .

As was noted above, in our case of the AdS/CFT correspondence, the bulk
field $\mathcal{V}_{M}$ corresponds to the boundary current $j^{\mu }$ of
the spin-3/2 field. In the field theory the S-matrix element in momentum
space is written in the form
\begin{equation*}
S_{fi}=\int d^{4}p^{\prime }d^{4}p\,j^{\mu }\varepsilon _{\mu }.
\end{equation*}
In our case, an expression corresponding to the current $j^{\mu }$ can be
extracted from (\ref{50}):
\begin{equation}
j^{\mu }=\overline{u}_{\alpha }\left( p\prime \right) \tilde{O}^{\alpha \mu
\beta }u_{\beta }\left( p\right) .  \label{52}
\end{equation}
Now we identify the boundary R-S field with the $\Delta $ baryons. Then
matrix element (\ref{50}) will describe $\gamma ^{\ast }\Delta \Delta $
interaction vertex and the matrix element (\ref{52}) will be identified with
the matrix element for $\Delta $ baryon current (\ref{23}). This suggestion
leads to identification of tensor $\tilde{O}^{\alpha \mu \beta }$ in (\ref
{52}) with the $O^{\alpha \mu \beta }$ in (\ref{24}). The terms of $\tilde{O}
^{\alpha \mu \beta }$, which have the same tensorial structure as those of $
O^{\alpha \mu \beta }$ will be corresponded to each other and integrals over
the fifth coordinate $z$ in the summands $\tilde{O}_{s}^{\alpha \mu \beta }$
will be identified with coefficients $a_{i}$ or $c_{i}$ of corresponding
term in $O^{\alpha \mu \beta }$. Thus, the coefficients $a_{i}$ and $c_{i}$
are expressed as following integrals of profile function $V\left( qz\right) $
and $F_{iL,R}\left( pz\right) $ over $z$ variable:

\begin{equation*}
a_{1}^{\left( i\right) }=\sqrt{2}\int\limits_{0}^{z_{m}}dz\ \;V\left(
qz\right) \left\{ Q_{i}\left[ z^{-2}+\left( 4m^{2}-q^{2}\right) k_{9}\right] 
\left[ F_{1L}^{\ast }\left( p\prime z\right) F_{1L}\left( pz\right)
+F_{2L}^{\ast }\left( p\prime z\right) F_{2L}\left( pz\right) \right] \right.
\end{equation*}
\begin{equation*}
\left. +\left[ -z^{-1}m\left( \eta ^{(s)}+\eta ^{(v)}\mathcal{T}
_{i}^{(3)}\right) -\left( k_{5}z^{-3}+2mk_{8}z^{-2}\right) mQ_{i}\left( 1-
\frac{q^{2}}{4m^{2}}\right) +k_{11}Q_{i}q^{2}z^{-1}+\right. \right.
\end{equation*}
\begin{equation*}
\left. \left. +Q_{i}\left( -k_{4}+z^{-2}m\left( 1-\frac{q^{2}}{4m^{2}}
\right) k_{7}-\frac{2q^{2}}{m}k_{12}\right) \right] \left[ F_{1L}^{\ast
}\left( p\prime z\right) F_{2L}\left( pz\right) +\right. \right.
\end{equation*}
\begin{equation*}
\left. \left. +F_{2L}^{\ast }\left( p\prime z\right) F_{1L}\left( pz\right)
\right] +6k_{13}Q_{i}\left[ F_{1L}^{\ast }\left( p\prime z\right)
F_{1L}\left( pz\right) -F_{2L}^{\ast }\left( p\prime z\right) F_{2L}\left(
pz\right) \right] \right\}
\end{equation*}
\begin{equation*}
+\sqrt{2}\int\limits_{0}^{z_{m}}dz\ \;\partial _{z}V\left( qz\right) \left\{ 
\left[ z^{-1}m\left( \eta ^{(s)}+\eta ^{(v)}\mathcal{T}_{i}^{(3)}\right) -
\frac{1}{2}k_{10}Q_{i}z^{-2}-2q^{2}Q_{i}\left( \frac{1}{m}
k_{12}+2k_{18}\right) \right] \right.
\end{equation*}
\begin{equation*}
\left. \times \left[ F_{1L}^{\ast }\left( p\prime z\right) F_{1L}\left(
pz\right) -F_{2L}^{\ast }\left( p\prime z\right) F_{2L}\left( pz\right) 
\right] +\frac{2q^{2}}{m}k_{18}Q_{i}\left[ F_{1L}^{\ast }\left( p\prime
z\right) F_{2L}\left( pz\right) +F_{2L}^{\ast }\left( p\prime z\right)
F_{1L}\left( pz\right) \right. \right.
\end{equation*}
\begin{equation}
\left. \left. -\partial _{z}F_{1L}^{\ast }\left( p\prime z\right)
F_{2L}\left( pz\right) -\partial _{z}F_{2L}^{\ast }\left( p\prime z\right)
F_{1L}\left( pz\right) -F_{1L}^{\ast }\left( p\prime z\right) \partial
_{z}F_{2L}\left( pz\right) -F_{2L}^{\ast }\left( p\prime z\right) \partial
_{z}F_{1L}\left( pz\right) \right] \right\}  \label{53}
\end{equation}
\begin{equation*}
a_{2}^{\left( i\right) }=\sqrt{2}m\int\limits_{0}^{z_{m}}dz\frac{1}{z^{3}}\
V\left( qz\right) \left\{ \left[ Q_{i}\left( k_{2}+k_{5}+k_{6}q^{2}+\left(
2mk_{8}+\frac{1}{2}k_{7}\right) z+k_{3}z^{2}\right) -\right. \right.
\end{equation*}
\begin{equation*}
\left. \left. -2\left( \eta ^{(s)}+\eta ^{(v)}\mathcal{T}_{i}^{(3)}\right)
z^{2}+8mQ_{i}z^{3}k_{12}\right] \left[ F_{1L}^{\ast }\left( p\prime z\right)
F_{2L}\left( pz\right) +F_{2L}^{\ast }\left( p\prime z\right) F_{1L}\left(
pz\right) \right] \right.
\end{equation*}
\begin{equation*}
\left. -4m^{2}Q_{i}k_{9}z^{3}\left[ F_{1L}^{\ast }\left( p\prime z\right)
F_{1L}\left( pz\right) +F_{2L}^{\ast }\left( p\prime z\right) F_{2L}\left(
pz\right) \right] \right\}
\end{equation*}
\begin{equation*}
+\sqrt{2}mk_{6}Q_{i}\int\limits_{0}^{z_{m}}dz\frac{1}{z^{3}}\partial
_{z}V\left( qz\right) \left[ \partial _{z}F_{1L}^{\ast }\left( p\prime
z\right) F_{2L}\left( pz\right) +\partial _{z}F_{2L}^{\ast }\left( p\prime
z\right) F_{1L}\left( pz\right) \right.
\end{equation*}
\begin{equation}
\left. +F_{1L}^{\ast }\left( p\prime z\right) \partial _{z}F_{2L}\left(
pz\right) +F_{2L}^{\ast }\left( p\prime z\right) \partial _{z}F_{1L}\left(
pz\right) \right]  \label{54}
\end{equation}
\begin{equation*}
c_{1}=-16m\sqrt{2}Q_{i}\int\limits_{0}^{z_{m}}dzV(qz)\left\{ \left[ k_{12}-
\frac{1}{8}k_{7}z^{-2}+\frac{m}{8}\left( \frac{1}{m}
k_{5}z^{-3}+2k_{8}z^{-2}+4z^{-1}k_{11}\right) \right] \right.
\end{equation*}
\begin{equation*}
\left. \times \left[ F_{1L}^{\ast }\left( p\prime z\right) F_{2L}\left(
pz\right) +F_{2L}^{\ast }\left( p\prime z\right) F_{1L}\left( pz\right) 
\right] +m\left( k_{18}+\frac{1}{4}z^{-2}k_{16}\right) \right.
\end{equation*}
\begin{equation*}
\left. \times \left[ \partial _{z}F_{1L}^{\ast }\left( p\prime z\right)
F_{1L}\left( pz\right) +F_{1L}^{\ast }\left( p\prime z\right) \partial
_{z}F_{1L}\left( pz\right) -F_{2L}^{\ast }\left( p\prime z\right) \partial
_{z}F_{2L}\left( pz\right) -\partial _{z}F_{2L}^{\ast }\left( p\prime
z\right) F_{2L}\left( pz\right) \right] \right.
\end{equation*}
\begin{equation*}
\left. -\frac{1}{8}m\left( z^{-2}k_{15}-5z^{-1}k_{17}\right) \left[
F_{1L}^{\ast }\left( p\prime z\right) F_{1L}\left( pz\right) +F_{2L}^{\ast
}\left( p\prime z\right) F_{2L}\left( pz\right) \right] \right\}
\end{equation*}
\begin{equation*}
-16m\sqrt{2}Q_{i}\int\limits_{0}^{z_{m}}dz\partial _{z}V(qz)\left\{ k_{18}
\left[ F_{1L}^{\ast }\left( p\prime z\right) F_{2L}\left( pz\right)
+F_{2L}^{\ast }\left( p\prime z\right) F_{1L}\left( pz\right) \right. \right.
\end{equation*}
\begin{equation*}
\left. -\partial _{z}F_{1L}^{\ast }\left( p\prime z\right) F_{2L}\left(
pz\right) -\partial _{z}F_{2L}^{\ast }\left( p\prime z\right) F_{1L}\left(
pz\right) -F_{1L}^{\ast }\left( p\prime z\right) \partial _{z}F_{2L}\left(
pz\right) -F_{2L}^{\ast }\left( p\prime z\right) \partial _{z}F_{1L}\left(
pz\right) \right]
\end{equation*}
\begin{equation}
\left. +\left( k_{12}-\frac{5}{2}k_{18}+\frac{1}{4}mk_{16}z^{-2}\right) 
\left[ F_{1L}^{\ast }\left( p\prime z\right) F_{1L}\left( pz\right)
-F_{2L}^{\ast }\left( p\prime z\right) F_{2L}\left( pz\right) \right]
\right\}  \label{55}
\end{equation}
\begin{equation}
c_{2}^{\left( i\right) }=8\sqrt{2}m^{3}Q_{i}k_{14}\int\limits_{0}^{z_{m}}dz\
\;z^{-3}V\left( qz\right) \left[ F_{1L}^{\ast }\left( p\prime z\right)
F_{1L}\left( pz\right) +F_{2L}^{\ast }\left( p\prime z\right) F_{2L}\left(
pz\right) \right]  \label{56}
\end{equation}

For brevity of these expressions we have used relations $F_{1R}\left(
pz\right) =-F_{2L}\left( pz\right) ,$ $F_{2R}\left( pz\right) =F_{1L}\left(
pz\right) $ and have wrote all coefficients in terms of $F_{1,2L}\left(
pz\right) $. Recall that the superscript $\left( i\right) $ shows a kind of $
\Delta $ baryons. The obtained here integral expressions for $a_{1,2}^{(i)}$
and $c_{1,2}^{(i)}$ can be applied for numerical studies of form factors ( 
\ref{25 a}) - (\ref{25 d}) and related physical quantities in the framework
of the hard-wall model AdS/QCD.

\section{Numerical analysis}
\label{sec4}

\begin{figure}[!h]
\begin{center}
\includegraphics[width=9cm]{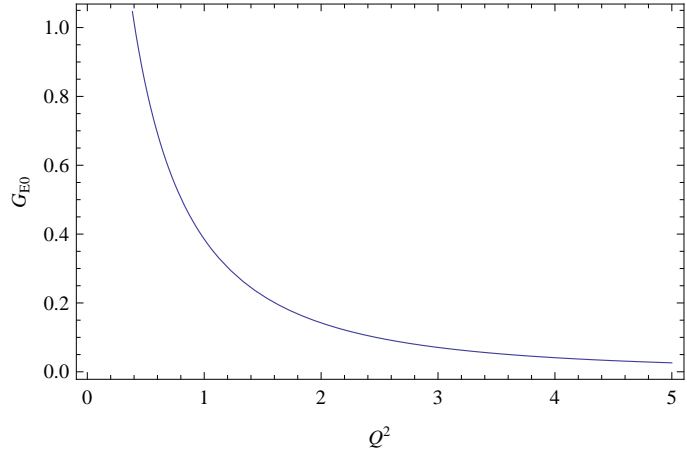}
\end{center}
\caption{ Charge form factor of $\Delta^{+}$ baryon}
\label{f2MB}
\end{figure}

In order to make a comparison with the electromagnetic form factors of $
\Delta $ baryons obtained in the field theory the
initial and final R-S fields should be taken on mass shell and the momenta $p$ and $p^{\prime }$ in $
F_{iL,R}$ in (\ref{53}) - (\ref{56}) should be set $p=p^{\prime }=m$. The time-like
region of $Q^{2}=-q^{2}$ should be considered. In this region the bulk to boundary propagator(\ref{15})
becomes
\begin{equation}
V\left( Q,z\right) =\frac{\pi }{2}Qz\left[ \frac{K_{0}\left( Qz_{m}\right) }{
I_{0}\left( Qz_{m}\right) }I_{1}\left( Qz\right) +K_{1}\left( Qz\right) 
\right]   \label{57}
\end{equation}
and its derivative equals to the following expression:
\begin{equation}
\partial _{z}V\left( Q,z\right) =\frac{\pi }{2}Q^{2}z\left[ \frac{
K_{0}\left( Qz_{m}\right) }{I_{0}\left( Qz_{m}\right) }I_{0}\left( Qz\right)
-K_{0}\left( Qz\right) \right] .  \label{58}
\end{equation}
As an example of application of (\ref{53}) - (\ref{56}) let us consider the
charge form factor $G_{E0}$ for one of the charged $\Delta $ baryons, i.e. for $\
\Delta ^{+}$. For this baryon in (\ref{53}) - (\ref{56}) we should take $Q_{2}=1$ and $\mathcal{T}
_{2}^{(3)}=1$ and then $G_{E0}$ will be written in terms of $Q^{2}$\
 as below:
\begin{equation*}
G_{E0}^{\left( 2\right) }\left( Q^{2}\right) =\left( 1+\frac{2}{3}\frac{Q^{2}
}{\left( 2m\right) ^{2}}\right) \left[ a_{1}^{\left( 2\right) }+\left( 1+
\frac{Q^{2}}{\left( 2m\right) ^{2}}\right) a_{2}^{\left( 2\right) }\right] -
\end{equation*}
\begin{equation}
-\frac{1}{3}\frac{Q^{2}}{\left( 2m\right) ^{2}}\left( 1+\frac{Q^{2}}{\left(
2m\right) ^{2}}\right) \left[ c_{1}^{\left( 2\right) }+\left( 1+\frac{Q^{2}}{
\left( 2m\right) ^{2}}\right) c_{2}^{\left( 2\right) }\right] .  \label{59}
\end{equation}
The coefficient functions $a_{1,2}^{(2)}$ and $c_{1,2}^{(2)}$ are reduced to
integrals of products of Bessel functions $J_{i}\left(
mz\right) $:
\begin{equation*}
A_{n}^{\pm }=\sqrt{2}C^{2}\int\limits_{0}^{z_{m}}dz\ \;V\left( Qz\right)
z^{n}\left[ J_{2}^{2}\left( mz\right) \pm J_{3}^{2}\left( mz\right) \right]
\end{equation*}
\begin{equation*}
B_{n}=\sqrt{2}C^{2}\int\limits_{0}^{z_{m}}dz\ \;V\left( Qz\right)
z^{n}J_{2}\left( mz\right) J_{3}\left( mz\right)
\end{equation*}
\begin{equation*}
C_{n}=\sqrt{2}C^{2}\int\limits_{0}^{z_{m}}dz\ V\left( Qz\right)
z^{n}J_{1}\left( mz\right) J_{2}\left( mz\right)
\end{equation*}
\begin{equation*}
D_{n}=\sqrt{2}C^{2}\int\limits_{0}^{z_{m}}dz\ \;\partial _{z}V\left(
Qz\right) z^{n}\left[ J_{2}^{2}\left( mz\right) -J_{3}^{2}\left( mz\right) 
\right]
\end{equation*}
\begin{equation*}
E_{n}=\sqrt{2}C^{2}\int\limits_{0}^{z_{m}}dz\partial _{z}V(Qz)z^{n}\left[
J_{1}\left( mz\right) J_{3}\left( z\right) +J_{2}^{2}\left( mz\right) \right]
\end{equation*}
\begin{equation*}
F_{n}=\sqrt{2}C^{2}\int\limits_{0}^{z_{m}}dz\partial
_{z}V(Qz)z^{n}J_{2}\left( mz\right) J_{3}\left( z\right)
\end{equation*}
and then in terms of these integrals the coefficient functions become equal to:

\begin{equation*}
a_{1}^{\left( 2\right) }=A_{3}^{+}+\left( 4m^{2}+Q^{2}\right)
k_{9}A_{5}^{+}+2m\left( 1+\frac{Q^{2}}{4m^{2}}\right) k_{5}B_{2}-2m\left( 1+
\frac{Q^{2}}{4m^{2}}\right) \left( k_{7}-2mk_{8}\right) B_{3}
\end{equation*}
\begin{equation*}
+\left[ 2m\left( \eta ^{(s)}+\eta ^{(v)}\right) +2k_{11}Q^{2}\right]
B_{4}+2\left( k_{4}+\frac{2Q^{2}}{m}k_{18}-\frac{2Q^{2}}{m}k_{12}\right)
B_{5}+6k_{13}A_{5}^{-}+
\end{equation*}
\begin{equation}
-\frac{1}{2}k_{10}D_{3}+m\left( \eta ^{(s)}+\eta ^{(v)}\right)
D_{4}+2Q^{2}\left( \frac{1}{m}k_{12}+2k_{18}\right) D_{5}+4Q^{2}k_{18}E_{5}
\label{60}
\end{equation}
\begin{equation*}
a_{2}^{\left( 2\right) }=-2m\left( k_{2}+k_{5}-k_{6}Q^{2}\right)
B_{2}-2m\left( 2mk_{8}+\frac{1}{2}k_{7}\right) B_{3}-2m\left[ k_{3}-2\left(
\eta ^{(s)}+\eta ^{(v)}\right) \right] B_{4}
\end{equation*}
\begin{equation}
-16mk_{12}B_{5}-4m^{3}k_{9}A_{5}^{+}-2m^{2}k_{6}E_{2}  \label{61}
\end{equation}
\begin{equation*}
c_{1}^{\left( 2\right) }=4mk_{5}B_{2}-4m\left(
k_{7}-2mk_{8}-2m^{2}k_{16}\right) B_{3}+16m^{2}k_{11}B_{4}+32m\left(
k_{12}+m^{2}k_{18}\right) B_{5}
\end{equation*}
\begin{equation*}
-8m^{3}k_{16}C_{3}-32m^{3}k_{18}C_{5}+2m^{2}\left( k_{15}-2k_{16}\right)
A_{3}^{+}-2m^{2}\left( 5k_{17}+8k_{18}\right) A_{4}^{+}
\end{equation*}
\begin{equation}
+32mk_{18}F_{5}-32m^{2}k_{18}E_{5}-8m^{2}k_{16}D_{3}-16m\left(
2k_{12}-5k_{18}\right) D_{5}  \label{62}
\end{equation}
\begin{equation}
c_{2}^{\left( 2\right) }=8m^{3}k_{14}A_{2}^{+}  \label{63}
\end{equation}
Main difficulty for numerical estimations is the absence of values of
coefficients $k_{i}$. We choose all these constants equal to
0,00000001. The dependence of $G_{E0}$ on $Q^{2}$ is drown
in Fig.~\ref{f2MB}.
This dependence is widely studied in the lattice model for QCD in 
\cite{16,17,27,28,29} and references therein. The behavior of $G_{E0}$ as a function of $
Q^{2}$ for $\Delta ^{+}$ baryon obtained here agrees with the one obtained in the
lattice QCD framework. 
Such qualitatively correct dependence of the form factor $G_{E0}$ on $Q^{2}$ indicates on 
rightness of inclusion into hard-wall model the interaction of electromagnetic field with 
the spin-3/2 fields. Similar numerical study can be done for other baryons
of this multiplet and for other form factors as well.

\section{Acknowledgements}

This work has been done under TUBITAK grant 2221 and author thanks this
organization. He thanks T.M.~Aliev for formulation of the problem and useful
discussions. The author also thanks K.O.~Ozansoy for inviting him to Ankara University and
hospitality during his visit.

\end{document}